\documentclass[prl,reprint,superscriptaddress,nofootinbib,twocolumn,nolongbibliography]{revtex4-2}

\usepackage[linktoc=page,colorlinks,urlcolor=blue,citecolor=blue,linkcolor=blue]{hyperref}
\usepackage{amssymb,amsmath}
\usepackage{graphicx}
\usepackage{natbib}
\usepackage{mathrsfs}
\usepackage{overpic}
\usepackage{gensymb}
\usepackage{lineno}
\usepackage{color}
\usepackage{upgreek}
\usepackage[dvipsnames]{xcolor}
\usepackage[letterpaper,textwidth=7in,top=.75in,bottom=.75in]{geometry}
\usepackage{textcomp}

\usepackage{pdfpages}
\makeatletter
\AtBeginDocument{\let\LS@rot\@undefined}
\makeatother

\newcommand{\snl}{Sandia National Laboratories, Albuquerque, New Mexico 87185, USA}
\newcommand{\cint}{Center for Integrated Nanotechnologies, Sandia National Laboratories, Albuquerque, New Mexico 87123, USA}

\begin{document}

\title{High-Resolution Short-Circuit Fault Localization in a Multi-Layer Integrated Circuit using a Quantum Diamond Microscope}
\date{\today}

\author{P. Kehayias}
\email{pmkehay@sandia.gov}
\affiliation{\snl}

\author{J. Walraven}
\affiliation{\snl}

\author{A. L. Rodarte}
\affiliation{\snl}

\author{A. M. Mounce}
\affiliation{\cint}

\begin{abstract}
As integrated circuit (IC) geometry and packaging become more sophisticated with ongoing fabrication and design innovations, the electrical engineering community needs increasingly-powerful failure analysis (FA) methods to meet the growing troubleshooting challenges of multi-layer (with multiple metal layers) and multi-chip components. In this work, we investigate a new electronics FA method using a quantum diamond microscope (QDM) to image the magnetic fields from short-circuit faults. After quantifying the performance by detecting short-circuit faults in a multi-layer silicon die, we assess how a QDM would detect faults in a heterogeneously integrated (HI) die stack. This work establishes QDM-based magnetic imaging as  a  competitive technique for electronics FA, offering high spatial resolution, high sensitivity, and robust instrumentation. We anticipate these advantages to be especially useful for finding faults deep within chip-stack ICs with many metal layers, optically-opaque layers, or optically-scattering layers.
\end{abstract}
\maketitle

\section{Introduction}
Integrated circuit (IC) feature sizes continue to shrink in accordance with Moore’s Law, and it is important for electronics failure analysis (FA) techniques to keep up with new process technologies \cite{asmFAdeskRef, wagnerFAtechniques}.  FA techniques must be able to identify defects in devices with denser layouts, weaker signals, and multiple stacked dice \cite{mooresLaw3Db, mooresLaw3Da}.   The shrinking geometries of planar transistors (down to 26 nm), FinFETs (down to 5 nm), and gate-all-around (GAA) transistors (down to 3 nm), and the increasing complexity of ``More than Moore" devices  all pose new FA difficulties \cite{morethanMoore}. Furthermore, many existing FA methods struggle to find faults in the bottom layers of components made from multiple stacked dice.  Having accurate information of the number of faults in a device and their locations in two or three dimensions (including depth) helps identify the root cause of the IC production problem.  This motivates a constant pursuit of new FA approaches to keep up with the increasing challenge, with the goal of being able to locate weak  faults (with small signal amplitudes) deep inside a device with high spatial resolution and quick measurement time.

\begin{figure*}[ht]
\begin{center}
\begin{overpic}[width=0.95\textwidth]{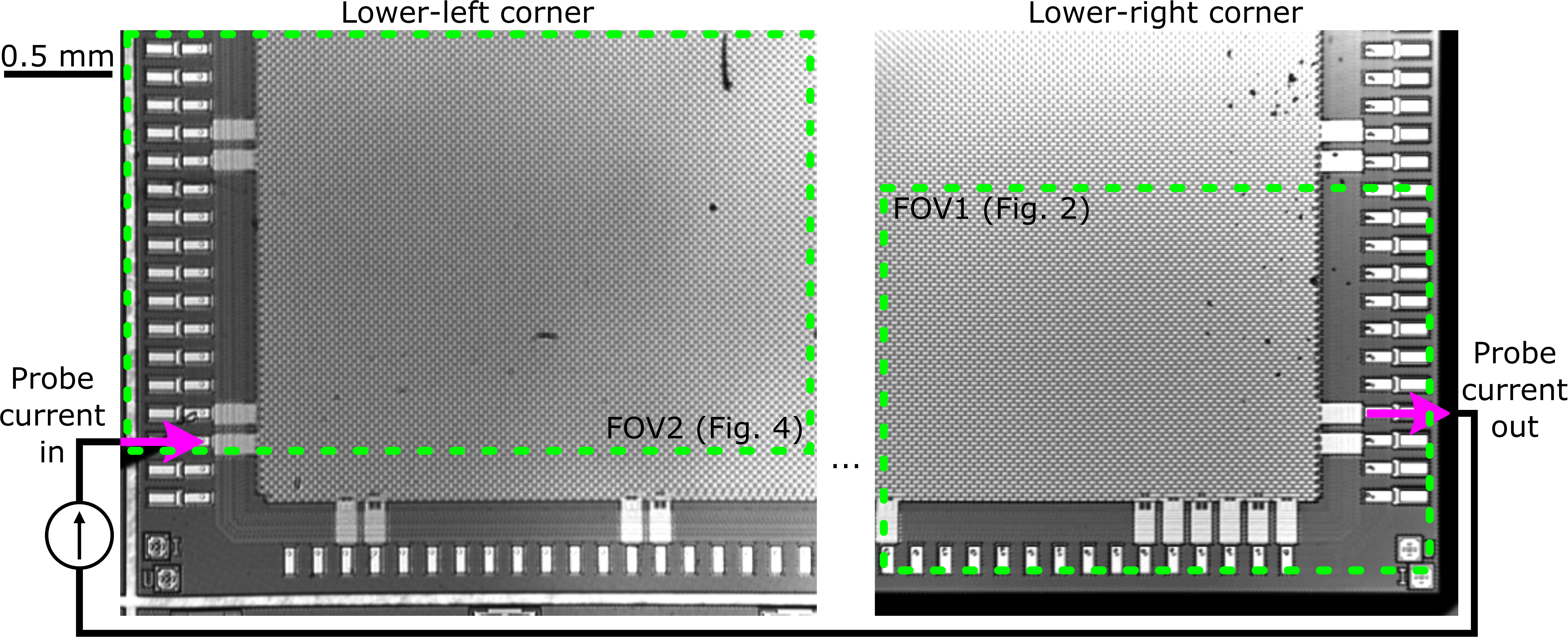}
\end{overpic}
\end{center}
\caption{\label{deviceOverviewB}
Photos of the lower-left and lower-right corners of the DUT, which we investigate in this work. We apply a probe current to a pad near the lower-left corner, after which it flows through the DUT and is collected from a pad near the lower-right corner. The green boxes indicate the fields of view (FOV) shown in later figures. 
}
\end{figure*}

In this paper we consider several leading electronics FA techniques, including scanning optical microscope approaches, thermal-based approaches, and magnetic imaging approaches \cite{asmFAdeskRef, wagnerFAtechniques}. With magnetic imaging, we measure the magnetic fields emitted from the internal electric current paths, which indicate where the current is flowing in the device, then use the magnetic field maps to locate short-circuit faults. Magnetic imaging is appealing because most IC materials (insulators, doped semiconductors, and metals) are transparent to magnetic fields, making it possible to look deep into an otherwise opaque device. In addition, high-performance magnetometers can sense weak magnetic fields, allowing us to detect and localize weak currents quickly.

When choosing a magnetic imaging technology for electronics FA, it is important to consider the magnetic sensitivity (noise floor), the signal-to-noise ratio (SNR), the spatial resolution, and the measurement standoff distance between the sensor and the magnetic sources. The SNR and the spatial resolution are closely related to the standoff distance; increasing the standoff distance generally leads to weaker magnetic fields and a coarser spatial resolution \cite{eduardoUpcont}. This suggests that an ideal magnetic imager should minimize the standoff distance to achieve the best SNR and spatial resolution. In addition, it is important to consider the technical details of the apparatus, including its operating conditions (e.g.~temperature), frequency range, and reliability. A standard tool for magnetic imaging is a scanning SQUID microscope \cite{hendersonFA}. Despite being a mature commercially-available instrument with a good magnetic noise floor, one weakness is that it scans a cryogenic sensor head across a room-temperature device, leading to a coarse spatial resolution ($\sim$50-100 $\upmu$m) and low reliability. 

Quantum magnetic sensing using nitrogen-vacancy (NV) centers in diamond is an emergent technique which enables high spatial resolution, high magnetic sensitivity, and integration with a variety of targets. One realization of this technology is the quantum diamond microscope (QDM) \cite{QDM1ggg, edlynQDMreview, tetienneQDMreview}, which circumvents the above drawbacks and is the focus of this work. By placing a synthetic diamond sample with a thin surface layer of magnetically-sensitive NV centers directly on top of a device, illuminating the NV layer with laser light, and imaging the NV fluorescence with a widefield optical microscope, we measure the magnetic field in each pixel using a camera. The QDM advantages include a standoff distance that can be made very small ($\sim$2 $\upmu$m air gap \cite{micromagnetPUFs}), a micron-scale spatial resolution, operation in ambient conditions, and nearly 100\% reliability. Earlier works showed how a QDM can image the magnetic fields from commercial ICs and other semiconductor devices \cite{qdmFPGA, NV555, NVAPAM, NVphotovoltaics}. Given that QDMs  already perform well when measuring working electronic devices, and the advantages listed above, this makes them promising for locating short-circuit faults in broken devices.

In this work, we used a QDM apparatus to locate short-circuit faults between two conducting layers in an application-specific integrated circuit (ASIC). We present six faults found in two dice, which were caused by copper defects between two conducting layers. To benchmark our technique against an established (thermal) FA method, we also used a thermally-induced voltage alteration (TIVA) imaging instrument to locate faults in the same two dice \cite{edcoleTIVAreview1, edcoleTIVAreview2}. TIVA is a standard electronics FA tool that uses laser heating to locate faults. While NV magnetic imaging and TIVA imaging differ in their technical maturity, operating principles, and limitations, both techniques detected short-circuit faults in the same locations, confirming that a QDM can perform FA.   Furthermore, we found that our QDM apparatus was able to detect faults with $35\times$ better SNR than the TIVA instrument, meaning that a QDM can find weaker faults considerably faster ($1200\times$ faster for the same SNR).  Our QDM instrument located two faults that were not seen with TIVA imaging, confirming the QDM SNR advantage. This work establishes the QDM as a high-performance apparatus for short-circuit fault localization, and we conclude with additional insights on how a QDM FA tool should work with heterogeneously integrated (HI) devices.

\begin{figure*}[ht]
\begin{center}
\begin{overpic}[width=0.9\textwidth]{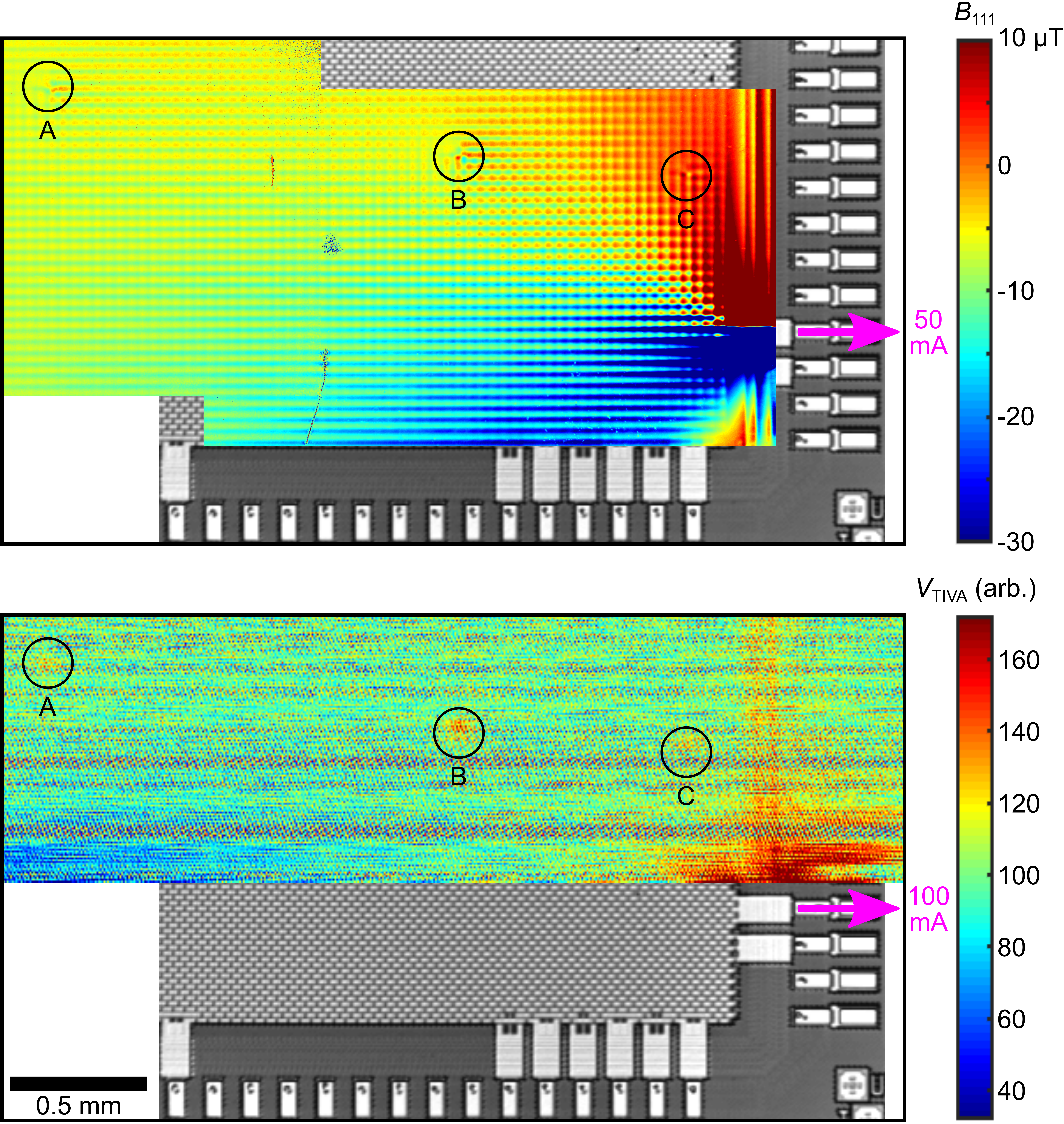}
\put(-2,96.5){\textsf{\Large a}}
\put(-2,45){\textsf{\Large b}}
\put(1,52){\includegraphics[width=0.10\textwidth]{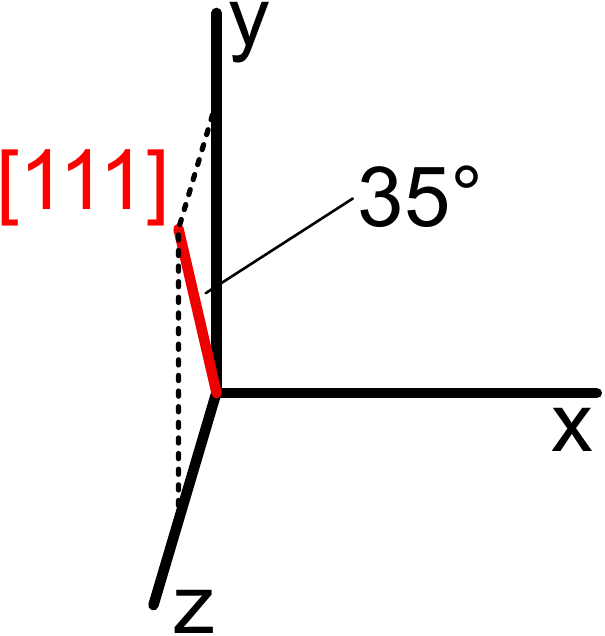}}
\end{overpic}
\end{center}
\caption{\label{QDMTIVAdie3}
(a) NV magnetic image with a 50 mA test current, with Faults A-C circled. The extra scratch marks are due to defects in the diamond surface. 
(b) TIVA image with a 100 mA test current, showing Faults A-C in the same locations.
}
\end{figure*}

\begin{figure*}[ht]
\begin{center}
\begin{overpic}[width=0.9\textwidth]{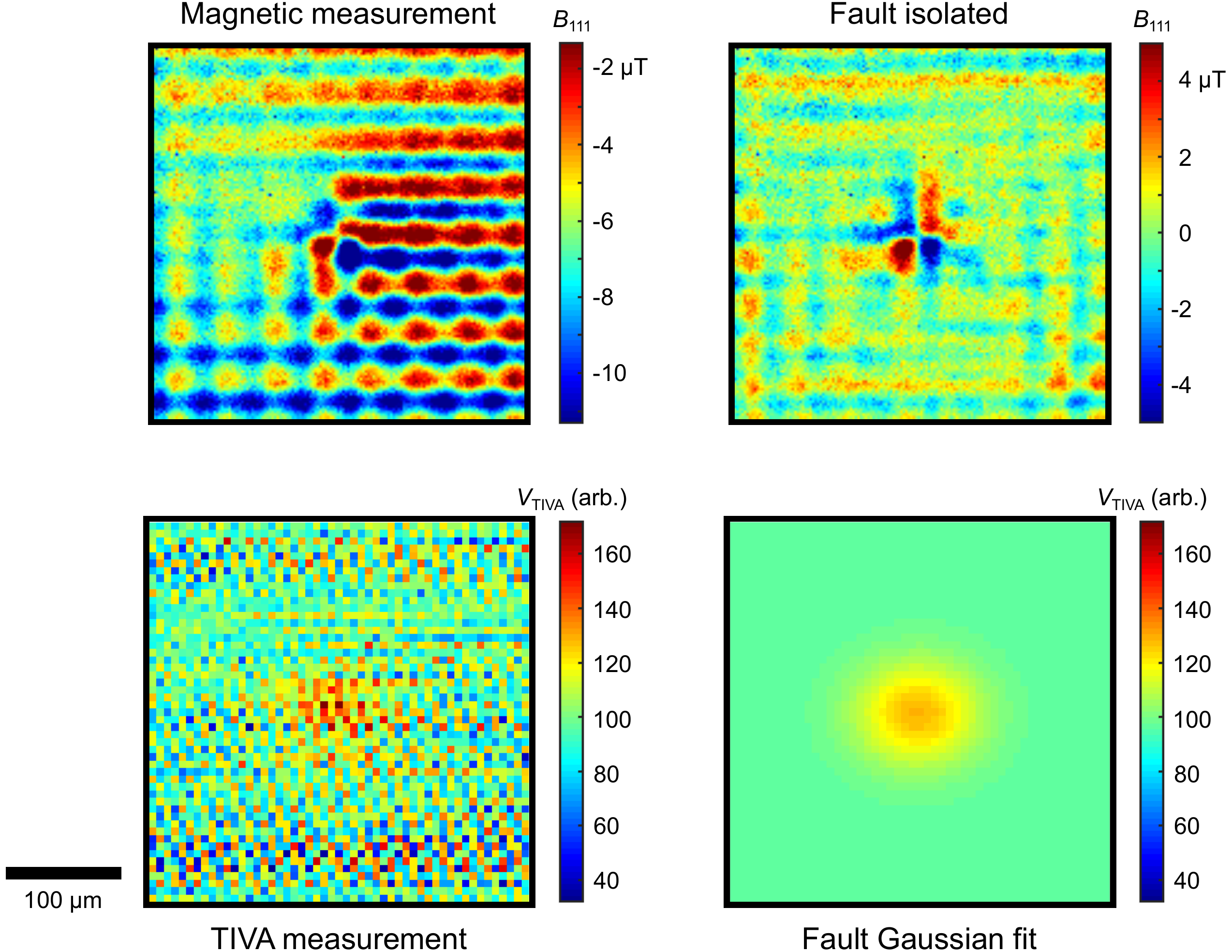}
\put(8,73){\textsf{\Large a}}
\put(55,73){\textsf{\Large b}}
\put(8,34){\textsf{\Large c}}
\put(55,34){\textsf{\Large d}}
\put(-2,40){\includegraphics[width=0.10\textwidth]{b111Coords.pdf}}
\end{overpic}
\end{center}
\caption{\label{compareSNR}
(a) QDM magnetic image for Fault B. (b) Magnetic image after suppressing the background magnetic fields, removing the grid pattern in the Fourier transform and removing a quadratic background field. From this image, we extract a $B_{\textrm{fault}} \approx 7.7~\upmu$T fault amplitude when applying a 50 mA test current. (c) TIVA image for Fault B. (d) 2D Gaussian fit for the TIVA image, from which we extract $V_{\textrm{fault}} = 32$ and $\delta V = 19$ for the fault amplitude and noise floor when applying a 100 mA test current. Note that both methods had a 20 minute experiment duration, but the TIVA imaging experiment had $2\times$ larger test current, $\sim$3$\times$3 larger pixels, and almost $4\times$ larger area.
}
\end{figure*}

\section{Results}

\subsection{Device under test}
Our device under test (DUT) is a silicon ASIC die \cite{suppl}. It is $27.45 \times 26.05$ mm$^2$ in size, and the outer perimeter is lined with pads. Internally, the DUT has multiple conducting layers, and was known from initial testing to have short-circuit faults  between the power planes in Layers 4 and 5 (2-10 $\ohm$ resistance between the relevant pads). However, the number of faults and their locations were not known \textit{a priori}. Our tasks are to test whether we can locate the short-circuit faults with a QDM setup, compare with the fault locations located with a TIVA imaging setup, and compare the performance between the two instruments.

Figure \ref{deviceOverviewB} shows optical images of the lower-left and lower-right corners of the DUT, where we searched for short-circuit faults. We present the results of two such dice. The fields of view being imaged are indicated as FOV1 and FOV2 for Die 1 and Die 2, respectively. We applied a test current between two pads using probe tips, then used NV magnetic imaging and TIVA imaging to search for short-circuit faults.

For the NV magnetic imaging experiments, we used a custom-made QDM apparatus with built-in probe tip manipulators to apply current to the relevant pads. The magnetic field images show the measured magnetic field component along the diamond [111] crystallographic direction ($B_{111}$), which is $\sim$35$\degree$ out of plane. For the TIVA measurements, we used a commercial system with a 1342 nm probe laser. Full details and schematics of the setups are included in the Supplementary Material \cite{suppl}.

\subsection{QDM imaging and TIVA imaging for FOV1}
Figure \ref{QDMTIVAdie3} shows two overlaid NV magnetic images and a TIVA image near the lower-right corner of a die (FOV1). We applied a 50 mA test current for the NV magnetic imaging experiments and a 100 mA test current for the TIVA imaging experiment. As the applied current disperses from the input pad and converges towards the output pad, it creates a grid pattern in the magnetic images similar to the grid pattern of Layers 4 and 5. The grid pattern contains defects caused by current traveling between Layers 4 and 5 along the $z$-axis, indicating short-circuit faults (labeled Faults A, B, and C). For comparison, the faults in the TIVA images appear as circular features (where laser heating at the fault location modifies the resistance) on top of a nominally-uniform background.

These initial results confirm that NV magnetic imaging can locate the same faults as TIVA imaging, a well-established FA technique. This is especially promising because the two methods use different phenomena to detect faults; NV magnetic imaging senses the internal currents in a device (both the intended and unintended current paths) while TIVA imaging detects changes in the DUT power demand caused by the increased resistance  from laser heating of metal defects in the current path, which may be responsible for short-circuit faults. To benchmark how NV magnetic imaging performs compared to TIVA imaging for this device, we now compare the SNR and the spatial resolution of each technique for Fault B.

\subsection{SNR assessment}
The QDM imaging SNR is given by
\begin{equation*}
\textrm{SNR}_{\textrm{QDM}} = \frac{B_{\textrm{fault}}}{\delta B}.
\end{equation*}
Here $B_{\textrm{fault}}$ is the magnetic field amplitude associated with a fault (normalized to a 1 A test current) and $\delta B$ is the standard deviation of the measured magnetic fields normalized to a 1 s measurement duration and a $1\times1~\upmu$m$^2$ pixel size. $\delta B$ is set by the photon shot noise of the NV fluorescence collected by the camera; it improves linearly with the pixel area and improves proportional to the square root of the measurement duration. We found that $\delta B = 30~\upmu$T. Although $\delta B$ was up to 4-8$\times$ better in our earlier works \cite{NV555}, here we reduced the laser power to minimize the risk of damaging the DUT or altering its internal currents due to stray light. Finally, if the current in the DUT is traveling through a resistive network, then $B_{\textrm{fault}}$ is proportional to the test current.

In Fig.~\ref{compareSNR}a-b, we evaluate $B_{\textrm{fault}}$ for Fault B. This is nontrivial to evaluate because we have to separate the $B_{\textrm{fault}}$ amplitude from the grid pattern superimposed on the fault, and both have a comparable amplitude. To separate them, we took the Fourier transform of Fig.~\ref{compareSNR}a and set the amplitudes of the frequency components from the grid pattern to zero. After calculating the inverse Fourier transform (with a suppressed grid pattern), we removed a quadratic background magnetic field. These fault isolation steps yield the magnetic field map of the isolated fault in Fig.~\ref{compareSNR}b. Fault B has a 7.7 $\upmu$T amplitude when using a 50 mA test current, which leads to $\textrm{SNR}_{\textrm{QDM}} = 5.1$ for a 1 A test current, 1 s measurement duration, and $1\times1~\upmu$m$^2$ pixel size.

Similarly, the TIVA imaging SNR is given by
\begin{equation*}
\textrm{SNR}_{\textrm{TIVA}} = \frac{V_{\textrm{fault}}}{\delta V}.
\end{equation*}
Here $V_{\textrm{fault}}$ is the TIVA voltage associated with a fault (normalized to a 1 A test current) and  $\delta V$ is the standard deviation of the measured voltages in the field of view normalized to a 1 s measurement duration and a $1\times1~\upmu$m$^2$ pixel size. Here we assume uncorrelated white noise, meaning that $\delta V$ improves linearly with the pixel area and improves proportional to the square root of the measurement duration. Furthermore, since TIVA imaging is a raster-scanning measurement, $\delta V$ scales with the square root of the image area for a given pixel size. We normalize the TIVA scan area to the QDM image area (1.31 mm $\times$ 2.09 mm  = 2.74 mm$^2$). Finally, $V_{\textrm{fault}}$ should likely scale with current. If the total resistance of the DUT ($R_0$) is proportional to that of a low-resistance fault between Layers 4 and 5, then laser heating should modify the resistance to $R_0 + \Delta R$, where $\Delta R = R_0 \alpha \Delta T$, $\alpha$ is a temperature coefficient, and $\Delta T$ is the temperature change. Under constant-current operation, $V_{\textrm{fault}} = \Delta V = I \Delta R$, so $V_{\textrm{fault}}$ should be proportional to the applied current and the probe laser power.

In Fig.~\ref{compareSNR}c-d, we evaluate $V_{\textrm{fault}}$ and $\delta V$ for Fault B. We fit the fault image with a 2D Gaussian function to find $V_{\textrm{fault}} = 32$ and $\delta V = 19$ in arbitrary units. Applying the relevant normalization factors, we get $\textrm{SNR}_{\textrm{TIVA}} = 0.15$ for 1 A test current, 1 s measurement duration, $1\times1~\upmu$m$^2$ pixel size, and 2.74 mm$^2$ image area.

Comparing $\textrm{SNR}_{\textrm{QDM}} = 5.1$ and $\textrm{SNR}_{\textrm{TIVA}} = 0.15$, we find that $\textrm{SNR}_{\textrm{QDM}}$ is $35\times$ larger for Fault B. This indicates a substantial sensitivity advantage when using NV magnetic imaging, allowing us to find weaker faults in less measurement time. Note that this SNR comparison may not be generalizable to all faults in the DUT. Although both techniques are sensitive to the current in the defect, $\textrm{SNR}_{\textrm{TIVA}}$ depends on other factors that do not affect $\textrm{SNR}_{\textrm{QDM}}$, such as how much nearby metal is contributing to the heat sinking.  Despite this detail, our SNR assessment gives us confidence that NV magnetic imaging can locate weaker faults than TIVA imaging. To confirm this finding, we studied a second die (FOV2 in Fig.~\ref{deviceOverviewB}) with weaker faults that were detectable with NV magnetic imaging but not with TIVA imaging in normal operating conditions, as described below. 

\begin{figure*}[ht]
\begin{center}
\begin{overpic}[width=0.90\textwidth]{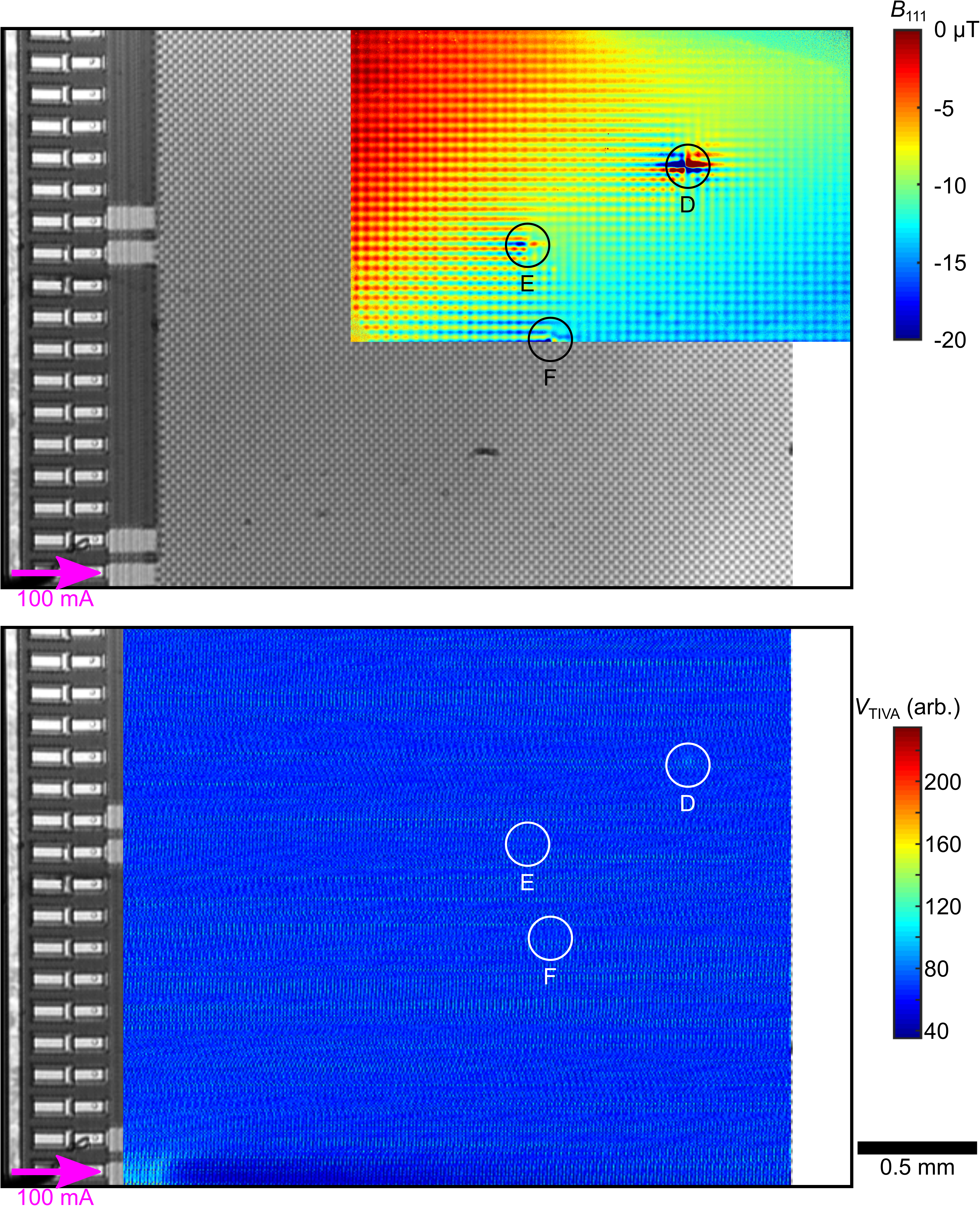}
\put(-2,97){\textsf{\Large a}}
\put(-2,46.5){\textsf{\Large b}}
\put(71.5,60){\includegraphics[width=0.10\textwidth]{b111Coords.pdf}}
\end{overpic}
\end{center}
\caption{\label{QDMTIVAdie4}
(a) QDM magnetic image for FOV2, showing Faults D-F.
(b) TIVA image for FOV2, showing the same three fault locations. Fault D is visible, but Faults E and F are too weak to see.
}
\end{figure*}

\begin{figure}[h]
\begin{center}
\begin{overpic}[width=0.45\textwidth]{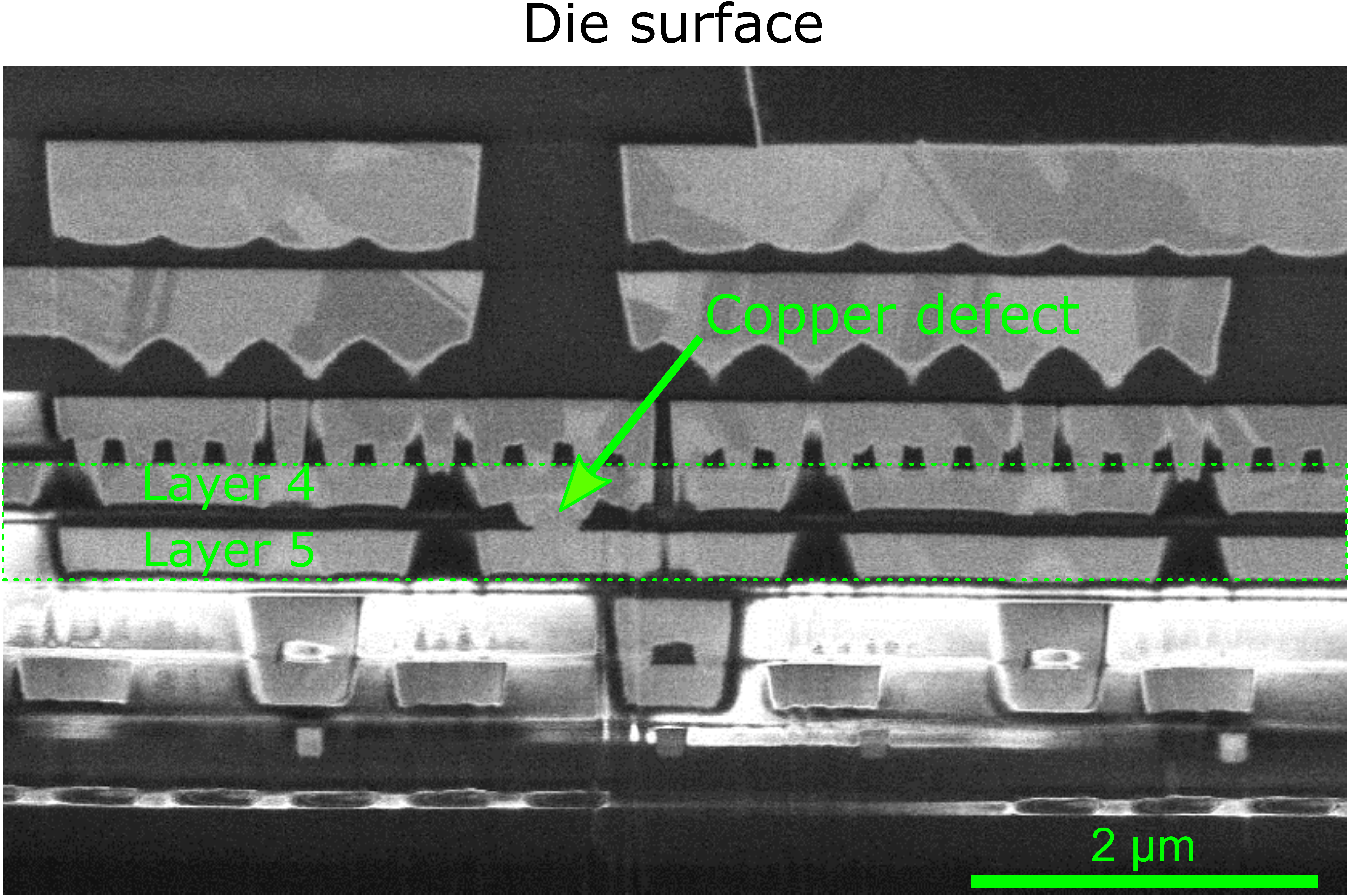}
\end{overpic}
\end{center}
\caption{\label{deviceCrossSec}
A scanning optical microscope (SEM) image of a die cross-section showing Fault F, which is caused by a copper defect between Layers 4 and 5. 
}
\end{figure}

\subsection{Spatial resolution comparison}
Spatial resolution has a variety of definitions, including the point spread function linewidth, the optical diffraction limit, or the closest distance before two features become indistinguishable. For this analysis we define spatial resolution as fault localization uncertainty, or how tightly we can localize an isolated fault in the $\{x,y\}$ plane, which is the spatial resolution metric relevant to electronics FA.

For NV magnetic imaging, Fault B has a location uncertainty of a few microns, indicated by the width of the cross shape in Fig.~\ref{compareSNR}b. Note that the $\{x,y\}$ location of a fault within the $8~\upmu\textrm{m} \times 8~\upmu\textrm{m}$ overlap between the traces in Layers 4 and 5 cause very different magnetic images. Simulating this phenomenon and comparing to the experimental magnetic images can yield a fault position uncertainty significantly smaller than the $16~\upmu$m periodicity for these conducting layers. For the TIVA image, the 2D Gaussian in Fig.~\ref{compareSNR}d has a standard deviation of 34 $\upmu$m. This standard deviation provides an estimate for the spatial resolution, though the centroid uncertainty can be made arbitrarily small with sufficiently-good SNR. Overall, we conclude that our QDM spatial resolution is $\sim$10$\times$ better than the TIVA imaging spatial resolution, though both could be enhanced with additional hardware improvements and post-processing efforts.

The spatial resolution for both techniques is limited by the optical diffraction limit $d = \frac{\lambda}{2 \textrm{NA}}$, where $\lambda$ is the fluorescence/illumination wavelength for the QDM/TIVA instrument and NA is the microscope objective numerical aperture. We get $d_{\mathrm{QDM}} = 1.4~\upmu$m for our QDM apparatus ($\lambda = 700$ nm, NA = 0.25) and $d_{\mathrm{TIVA}} = 4.8~\upmu$m for our TIVA apparatus ($\lambda = 1342$ nm, NA = 0.14). Given the same NA, $d_{\mathrm{QDM}}$ should be $1.9\times$ better than $d_{\mathrm{TIVA}}$. Both could be improved with a better NA, though NV magnetic imaging can also employ super-resolution microscopy or atomic force microscopy (AFM) techniques \cite{chenNVsuperres, afm3}. In practice, the standoff distance usually limits the QDM spatial resolution, while heat spreading usually limits the TIVA spatial resolution \cite{paiTIVAthermalModeling}.

\subsection{QDM imaging and TIVA imaging for FOV2}
Figure \ref{QDMTIVAdie4} shows magnetic and TIVA images for a second die, in an area closer to the middle of the die where the currents are weaker. We applied a 100 mA test current for both the NV magnetic imaging and the TIVA imaging experiments, and we see three additional short-circuit faults, labeled Faults D, E, and F. Both techniques see a strong short-circuit fault (Fault D), though it is barely visible in the TIVA image. Furthermore, the NV magnetic image sees two additional faults (Faults E and F), which have magnetic field amplitudes that are $\sim$20-40$\times$ weaker than that of Fault D. Faults E and F are too weak to see in the TIVA image, confirming our earlier finding that $\textrm{SNR}_{\textrm{QDM}}$ is better than $\textrm{SNR}_{\textrm{TIVA}}$. Given knowledge of their locations from the magnetic image, we were later able to locate them after several SNR enhancements to the TIVA measurement \cite{suppl}.

After finishing the QDM and TIVA imaging experiments, we used a focused ion beam (FIB) to cross-section the DUT at the fault locations, then used a scanning electron microscope (SEM) to image the cross-sections. Figure \ref{deviceCrossSec} includes a cross-section for Fault F, showing the copper defect causing a short between the conducting layers.

\begin{figure*}[ht]
\begin{center}
\begin{overpic}[width=0.95\textwidth]{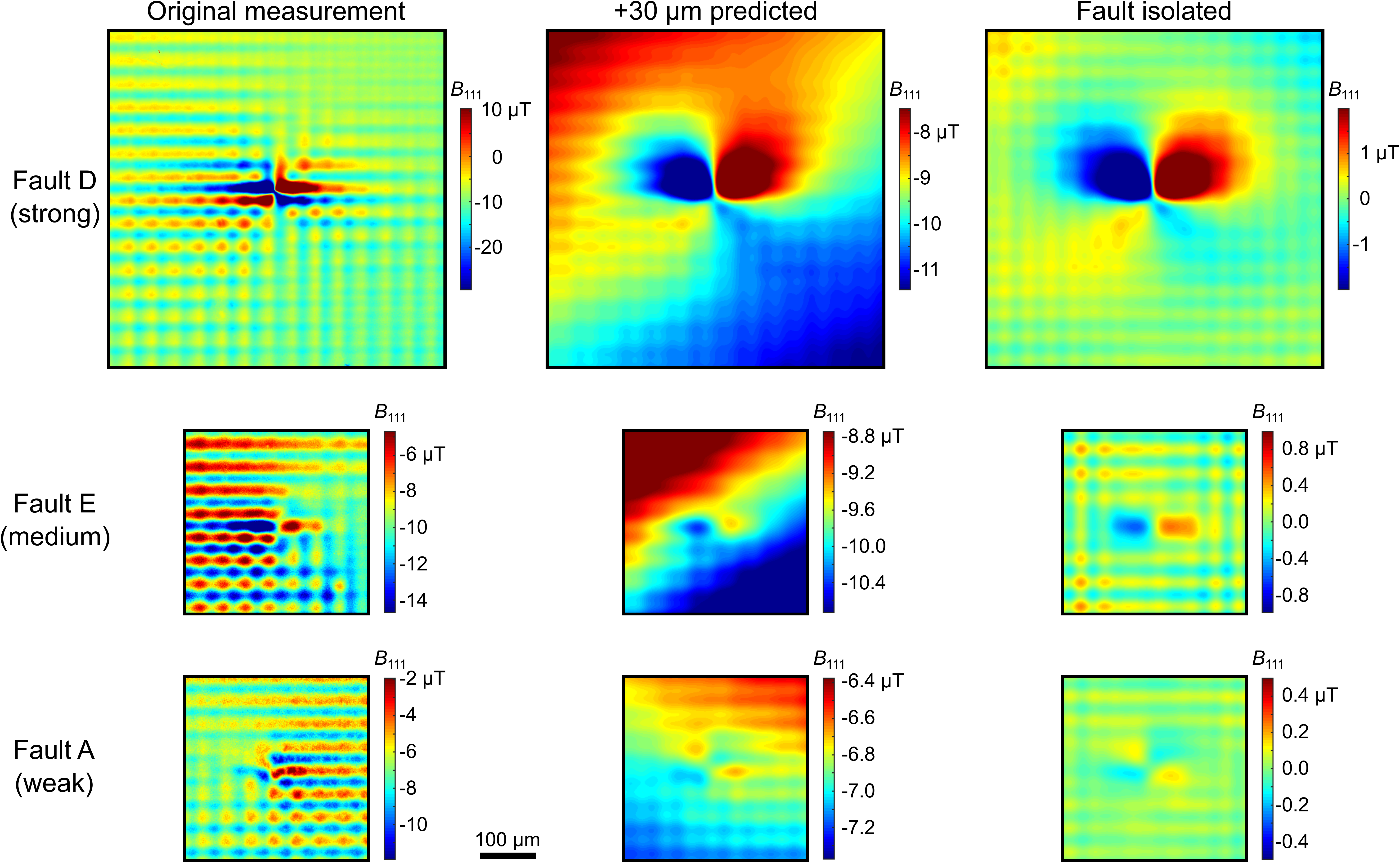}
\put(31.5,3){\includegraphics[width=0.10\textwidth]{b111Coords.pdf}}
\end{overpic}
\end{center}
\caption{\label{upcont}
Magnetic images for Faults D, E, and A, representing strong, medium, and weak-amplitude faults. Using the measured magnetic images (left column), we simulate the magnetic images if the die was part of an HI device, with an additional 30 $\upmu$m die on top (middle column). We applied the same fault isolation post-processing to remove the background ripples and gradients as in Fig.~\ref{compareSNR} (right column).
}
\end{figure*}

\section{Magnetic imaging for an HI device}
Having established NV magnetic imaging as a promising electronics FA approach for locating short-circuit faults, we now consider how this approach can be used for a broader range of devices. Perhaps the most challenging devices for NV magnetic imaging to measure are those with weak currents in deep layers (i.e.~ICs with many conducting layers and stacked dice), since the magnetic field amplitude and spatial resolution will degrade with the increased standoff distance. 

The DUTs imaged in Fig.~\ref{QDMTIVAdie3} and Fig.~\ref{QDMTIVAdie4} are intended to be part of an HI device, with an additional 30 $\upmu$m die to be mounted on top. To simulate the magnetic images for short-circuit faults if they were measured in an HI device, we calculated the expected magnetic field images with 30 $\upmu$m of additional standoff distance \cite{eduardoUpcont}, shown in Fig.~\ref{upcont}. We evaluated the anticipated performance for three typical fault magnetic amplitudes: strong (Fault D), medium (Fault E), and weak (Fault A).  As expected, the grid patterns and fault features become weaker and coarser, but fortunately the fault features are still visible with the added standoff distance, especially after using the fault isolation steps described above. This suggests that NV magnetic microscopy could also work well for HI devices, despite the increased standoff distance. However, since weak magnetic features become more difficult to isolate from the background magnetic field with increasing standoff distance,  minimizing the measurement standoff distance is always preferable. Despite these limitations, Fault D (which has a large $B_{\textrm{fault}} \approx$ 120 $\upmu$T amplitude and a small background gradient) should be visible as far as $\sim$150 $\upmu$m away. This suggests that magnetic imaging (using a QDM or other methods) could still be viable for finding short-circuit faults deep within HI devices, stacked-die ICs, devices using backside power delivery, and other devices with limited optical access to the deep layers.

\section{Discussion}
Magnetic imaging and TIVA imaging have different strengths and trade-offs when used to locate short-circuit faults. With magnetic imaging, the magnetic field amplitude and spatial resolution get worse with larger standoff distance, but the magnetic signal should be unimpeded by any materials (metals, semiconductors, or insulators) between the fault and the sensor. For TIVA imaging, the materials  matter; it is difficult to find a fault surrounded by metal due to to its opacity and thermal conductivity. However, transparent semiconductor and insulator layers should not impede the performance significantly (e.g.~imaging conducting layers through back of the bulk silicon substrate). Despite its reduced sensitivity to deeper sources, magnetic imaging can still be an appealing choice for finding faults in multi-layer dice, stacked-die devices, and HI devices. In particular, if a fault is in a lower die under dense metallization with high interconnect density and underfill, magnetic imaging could be a viable FA option despite the performance sacrifice, while TIVA imaging might be impossible.

Compared to other magnetic imaging approaches, such as scanning SQUID microscopy, scanning giant magnetoresistance (GMR) microscopy, scanning magnetic tunnel junction (MTJ) microscopy, and magnetic force microscopy (MFM), NV magnetic imaging offers several advantages \cite{neoceraShorts, felt_SQUID_MR, MTJshorts, MFMshorts2}. Unlike these other techniques, a QDM acquires the magnetic information in all pixels simultaneously (rather than raster-scanning). This means a QDM has no moving parts when running, which is ideal when testing devices energized using probe tips. This also means that any drift during acquisition should be common to all pixels and easily removed, while drift during a raster-scanning experiment may be difficult to remove. Finally, a QDM can operate in ambient conditions, does not require a magnetic shield, has nearly 100\% reliability, and is inexpensive compared to other FA tools. These features make a QDM approach appealing to the electronics FA, power electronics, microelectronics manufacturing, very large scale integration (VLSI), and multi-chip module (MCM) communities.

For short-circuit faults deeper than $\sim$50-100 $\upmu$m, a scanning SQUID microscope will likely outperform a QDM for fault detection, though the QDM suitability for 2D, 2.5D, and 3D components depends on the overall device thickness. This is because for deep faults, we are unable to exploit the QDM advantage of having a few-micron standoff distance (the NV and SQUID standoff distances will be comparable), and a scanning SQUID microscope should have a better magnetic noise floor ($\sim$20 pT/$\sqrt{\mathrm{Hz}}$ ) \cite{wikswoReview}. However, a scanning SQUID microscope measuring the $B_z$ component of the magnetic field is insensitive to currents in the $z$ direction, while the vector magnetic field information from a QDM can be used to extract more insights about the internal currents within the devices being studied. Finally, any image analysis techniques used to enhance the spatial resolution of a coarse-resolution magnetic imager can also be applied to QDM images, with the added benefit of the pre-processed magnetic image having a closer standoff distance.

\section{Conclusions}
In this work, we investigated how an NV magnetic imaging apparatus can be used for electronics FA by locating short-circuit faults between conducting power planes in a multi-layer silicon ASIC device. Validating NV magnetic imaging with the industry-standard TIVA imaging technique, we found that NV magnetic imaging had a significantly better SNR ($35\times$ better for Fault B), meaning we can find weaker faults (e.g.~Faults E and F) in less time ($1200\times$ faster for the same SNR). We also showed that QDM imaging can be extended to HI devices and ICs made with stacked dice. As a result, we believe this work establishes the QDM as an effective instrument for short-circuit fault localization. We anticipate further exploration into how QDM vector magnetic imaging, sophisticated forward-model and inverse-model calculations, or magnetic image analysis using a machine learning algorithm may benefit fault localization \cite{hollenbergCurrentImg, fpgaBackThin, edlynIstfa2021}. In addition, using a QDM apparatus to image AC magnetic fields could open new electronics FA possibilities as well \cite{maletinskyMWimger}.

\section{Acknowledgements}
We thank Paiboon Tangyunyong and Edward Cole for their technical advice on the manuscript, and William Mook for his work in producing the focused ion beam cross-section images. Sandia National Laboratories is a multi-mission laboratory managed and operated by National Technology and Engineering Solutions of Sandia, LLC, a wholly owned subsidiary of Honeywell International, Inc., for the DOE's National Nuclear Security Administration under contract DE-NA0003525. This work was funded, in part, by the Laboratory Directed Research and Development Program and performed, in part, at the Center for Integrated Nanotechnologies, an Office of Science User Facility operated for the U.S.~Department of Energy (DOE) Office of Science. 


%

 \clearpage
 

\section*{Supplementary material}

# Supplemental Material for "High-Resolution Short-Circuit Fault Localization in a Multi-Layer Integrated Circuit using a Quantum Diamond Microscope"

P. Kehayias,[1, *] J. Walraven,[1] A. L. Rodarte,[1] and A. M. Mounce[2]

[1]*Sandia National Laboratories, Albuquerque, New Mexico 87185, USA*
[2]*Center for Integrated Nanotechnologies, Sandia National Laboratories, Albuquerque, New Mexico 87123, USA*

## EXPERIMENTAL DETAILS AND ENHANCED-SENSITIVITY TIVA RESULTS

### Device under test

Figure S1 shows a photo of an example die studied in this work. Die 1 and Die 2 were made on a silicon wafer, and we performed the experiments on the DUTs before wafer dicing. Internally, the DUT has multiple metal layers, and our work involved searching for short-circuit faults between the power planes in Layers 4 and 5. These layers contain orthogonal rows and columns arranged in a grid pattern, as shown in the Fig. S1 inset. We applied test currents through probe tips placed on the relevant pads.

### QDM apparatus

Our QDM setup combines an optical fluorescence microscope (used for magnetic imaging) and a probe station (used to apply test currents to the DUT), as shown in Fig. S2a. In the QDM setup, we place a $4 \times 4 \times 0.5$ mm$^3$ diamond sample (Sample A in Ref. [S1]) with a 4 μm sheet of magnetically-sensitive NV centers on top of the DUT (Fig. S2b). The NV electronic ground-state magnetic sublevels are sensitive to the magnetic field $B_{111}$ along the N-V axis (the [111] crystallographic direction), leading to the $m_s = 0$ to $m_s = \pm 1$ transition frequencies being shifted by $\pm \gamma B_{111}$, where $\gamma \approx 28$ GHz/T is the NV gyromagnetic ratio. We illuminate the NV layer using 532 nm laser light from the side of the microscope objective, which pumps the NVs into the $m_s = 0$ sublevel. Applying a resonant microwave field drives NVs from the $m_s = 0$ to the $m_s = \pm 1$ sublevels, reducing the NV fluorescence emission. Sweeping the microwave frequency and imaging the NV fluorescence with the microscope, we obtain an optically-detected magnetic resonance (ODMR) spectrum for each pixel in the microscope field of view. By fitting the ODMR spectra for each pixel to extract the center frequencies, we obtain a $B_{111}$ map detected by the NV layer [S2].

In this setup, we used a static 1.5 mT bias field along the N-V axis, and subtracted the magnetic field maps acquired using equal and opposite test currents applied to the device to remove any contributions to the magnetic images other than the test current (such as the bias magnetic field). We used 200 mW of laser power, and imaged a $1.31 \times 2.09$ mm$^2$ area with a 1.09 μm pixel size and a 20 minute typical experiment duration. Finally, to prevent the laser light and NV fluorescence from generating photocarriers in the DUT, we applied a protective coating (5 nm Ti, 150 nm Ag, and 150 nm Al$_2$O$_3$) to avoid altering the internal currents in the DUT with stray light [S3].

### TIVA apparatus

Thermally-induced voltage alteration (TIVA) is a technique for localizing defects in semiconductor ICs (Fig. S3) [S4, S5]. Illuminating the DUT with a Nd:YVO$_4$ continuous-wave monochromatic laser (1342 nm, 0.924 eV) with a photon energy below the silicon band gap (1.14 eV) causes localized heating, generating voltage gradients from thermal gradients (Seebeck effect) without creating electron-hole pairs or photocurrents. For TIVA imaging, localized heating from the focused laser beam alters the resistance of a short-circuit fault, producing a change in the IC power demand. Increased sensitivity is achieved when the DUT is under constant-current biasing. As we scan the laser beam across the DUT surface, we record the changes to the applied voltage and power demand, which indicates the heating (and fault) locations.

We used a laser-scanning microscope (Checkpoint InfraScan 400TDM) to map the thermally-induced voltage changes and the corresponding reflected laser intensities, allowing us to spatially localize the defect sites. For our measurements, we typically used 140 mW of 1342 nm laser light, a $3.21 \times 3.21$ mm$^2$ scan area with a 6.27 μm pixel size (512×512 pixels), and a 20 minute measurement duration. After finishing the scan, the instrument provides

images of the laser-induced voltage change normalized to an 8-bit grayscale image (0 to 255 in arbitrary units) and the reflected light intensity, which we use for the overlays in the main text.

## TIVA image of FOV2 with sensitivity enhancements

To check whether Faults E and F in FOV2 could feasibly be detected in a TIVA imaging experiment, we modified the TIVA imaging experiment to improve the noise floor. We used a 4× smaller scan area, 3× longer averaging time, and 4 × 4 pixel binning, totalling a ∼14× SNR enhancement. Figure S4 shows the resulting TIVA image. Despite these enhancements, Faults E and F are barely visible in this TIVA image. However, this confirms the fault positions are consistent with the NV magnetic image (Fig. 4a in the main text).

---

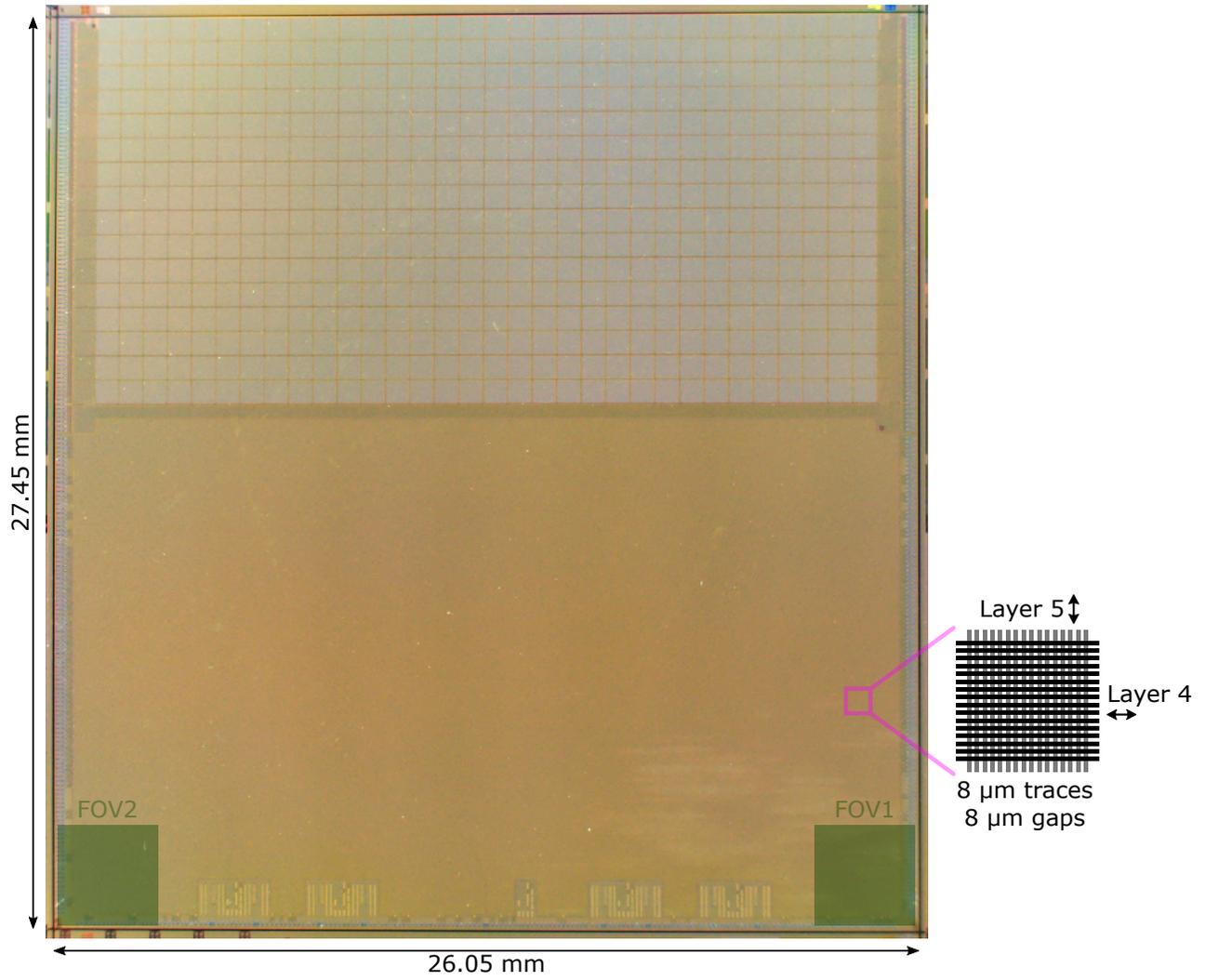

**Supplementary Figure** S1. Image of the device under test (DUT) used in this work, an ASIC with multiple conducting layers. The fields of view examined in this work (FOV1 and FOV2) are highlighted in green, and the inset shows the grid layout for the relevant conducting layers (Layers 4 and 5).

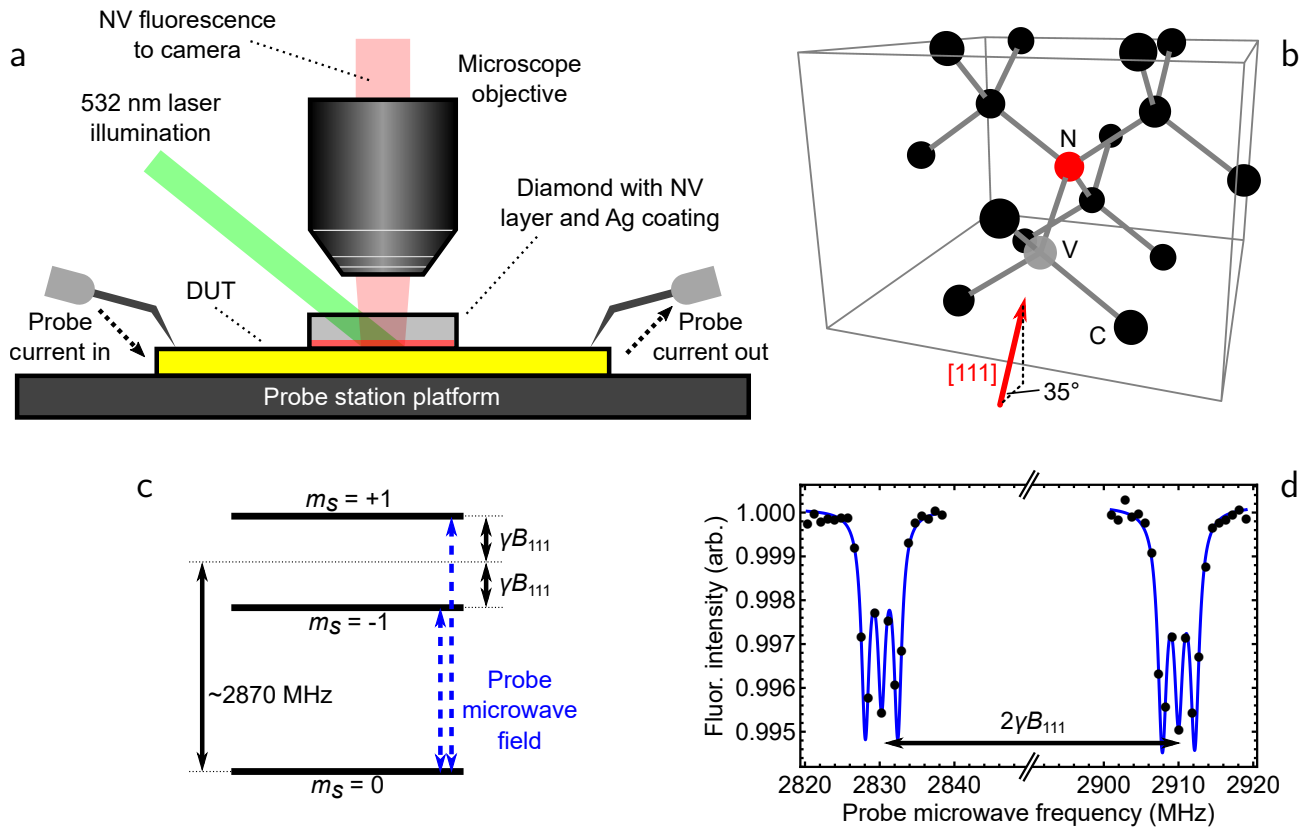

**Supplementary Figure** S2. (a) Schematic of the combined QDM/probe station setup for measuring a DUT, in this case an ASIC die. A diamond sample with a surface NV layer (red stripe) is placed on top of the DUT, illuminated with 532 nm laser light, and the NV fluorescence is imaged by the microscope. The microwave wire and bias $B_{111}$ magnetic field are not shown. (b) NV center in the diamond carbon lattice, with an arrow showing the [111] crystallographic direction. (c) Energy level diagram of the NV ground-state magnetic sublevels, indicating the $\pm\gamma B_{111}$ Zeeman effect and the zero-field splitting (~2870 MHz). (d) Example single-pixel ODMR spectrum; each lineshape is split into three peaks due to the $^{14}$N hyperfine interaction [S6].

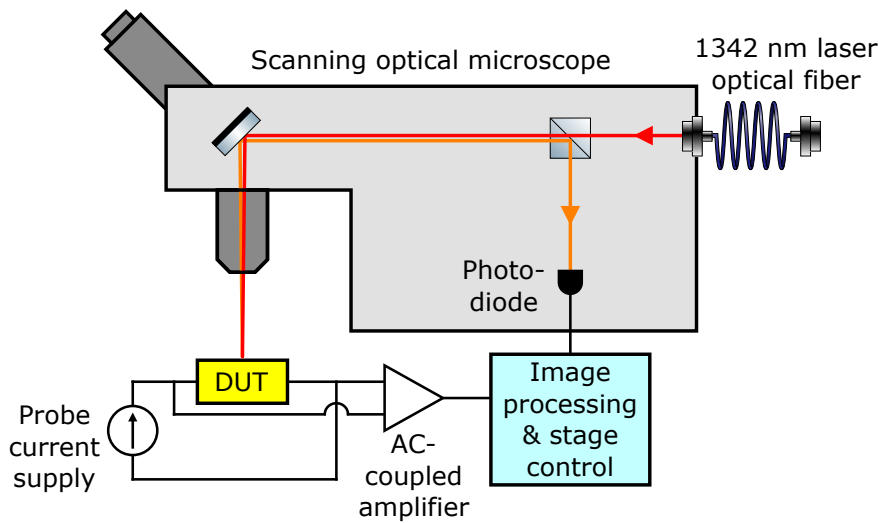

**Supplementary Figure** S3. Schematic of our TIVA setup. We illuminate the DUT with 1342 nm laser light (red beam), and we measure the reflected light (orange beam) with a photodiode. While applying a constant current to the DUT, the voltage across the DUT is also recorded. The scanning optical microscope apparatus moves across the DUT surface, imaging the reflected light intensity and TIVA voltage at each pixel.

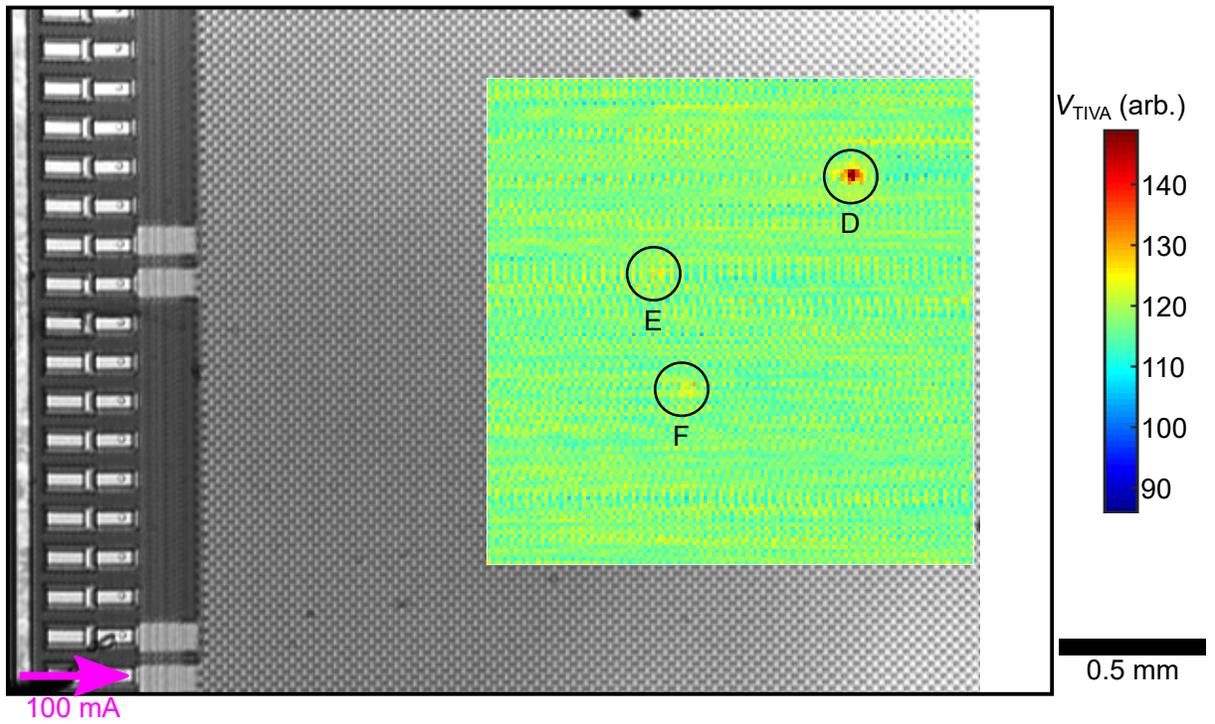

**Supplementary Figure** S4. TIVA image of FOV2 with an improved noise floor (4× smaller scan area, 3× longer averaging time, 4 × 4 pixel binning). Faults E and F (which are clearly visible in the NV magnetic image) are barely visible despite these improvements, showing that NV magnetic imaging is able to find weaker faults in less time than TIVA imaging.



\end{document}